\documentclass[showpacs,showkeys,preprintnumbers,amsmath,amssymb,twocolumn,prl,floatfix]{revtex4}

\usepackage{amssymb,amsfonts}
\usepackage{amsmath}
\usepackage{graphicx}
\graphicspath{{Figures/}}
\begin{document}

\title{Decoherence and Spin Echo  in Biological Systems}

\author{Alexander I. Nesterov}%
   \email{nesterov@cencar.udg.mx}
\affiliation{Departamento de F{\'\i}sica, CUCEI, Universidad de Guadalajara,
Av. Revoluci\'on 1500, Guadalajara, CP 44420, Jalisco, M\'exico}

\author{Gennady P.  Berman}
 \email{gpb@lanl.gov}
\affiliation{Theoretical Division, T-4, Los Alamos National Laboratory, and the New
Mexico Consortium,  Los Alamos, NM 87544, USA}

\date{\today}

\begin{abstract}
The spin echo approach is extended to include bio-complexes for which the interaction with dynamical noise is strong. Significant restoration of the free induction decay signal due to  homogeneous (decoherence) and inhomogeneous (dephasing) broadening  is demonstrated analytically and numerically, for both an individual   dimer of interacting chlorophylls and for an ensemble of dimers. This approach is based on an exact and closed system of ordinary differential equations that can be easily solved for a wide range of parameters that are relevant for bio-applications.
\end{abstract}



\preprint{LA-UR-15-20113}

\maketitle


 The femtosecond spectroscopic and measuring technologies, developed during last few decades, have allowed experimental scientists to examine very rapid dynamical processes, including  those in biological systems. In particular, it was demonstrated that, even at room temperature, photosynthetic bio-complexes exhibit collective quantum coherence (CQC) during primary electron transfer (ET) processes that occur on the time-scale of some hundreds of femtoseconds \cite{BOOK}. 
 
The CQC is resulted from the fact that the primary processes of exciton transfer and charge separation are so rapid (on a time-scale of a few picoseconds) that the  environment does not have time to recombine the exciton and  destroy the CQC. 

One of the methods for studying the CQC effects in bio-systems is the spin echo spectroscopy,  which was initially developed in the nuclear magnetic resonance \cite{AB}.   
 This spin echo procedure allows one to reduce the effects of the ``inhomogeneous broadening" (dephasing) and to  increase the time of the signal generated by the transverse magnetization (free induction, FI).  Recently, the spin echo technique has also been used to analyze quantum coherent effects in an ensemble of effective $1/2$ spins (chlorophyll-based dimers) in photosynthetic bio-complexes. (See \cite{FL1}, and references therein.)

During the last decade, it was recognized that the spin echo technique can be successfully applied not only to an ensemble of spins, but also to an individual spin, or to an individual two-level quantum system (TLS) \cite{ICJM}. In this case, inhomogeneous broadening is absent. But still one can partly suppress the broadening of the FI decay  resulted from a pure dynamical (time-dependent) noise.  In \cite{ICJM,BGA,GABS,NB1} the spin echo approach was applied to a single quantum two-level system - the superconducting qubit. In this case, the qubit is considered in the so-called diagonal representation, with the main characteristic being the energy gap (usually tens of gigahertz) between the ground and the excited states. The dynamical noise, acting on the qubit, is generated by the time-fluctuations of the electron charge, bias current, external magnetic flux, and other sources, and these all are  relatively weak. 

In contrast to a superconducting qubit, an individual effective two-level quantum system (a dimer) in a bio-complex is usually characterized in the so-called site representation, using such main parameters as the energy gap  (redox potential) between the excited states of the chlorophylls realizing this dimer, and the matrix element of the dipole-dipole or the exchange interactions between these two excited states.  Also, the dimer usually experiences a strong interaction with the protein environment, caused by dynamical noise (characterized by the reconstruction energy), that  must be taken into account in bio-applications of the spin echo technique.  

In this Letter, we analyze analytically and numerically the spin echo technique for both an individual dimer and for an ensemble of dimers in bio-complexes, for the case of strong interaction with the dynamical noise.  We show how both the dynamical and the inhomogeneous broadening can be suppressed by the spin echo pulses. Our conclusion is that even for strong dynamical noise, the spin echo approach  can serve as the useful complementary spectroscopic technique for characterizing the bio-complexes that include both individual dimers and an ensemble of dimers.

We consider a dimer  composed of the excited states of two chlorophylls, Chl1 and Chl2. We assume that each chlorophyll experiences a diagonal noise,  provided by the protein environment, which is described by the random variable, $\xi(t)$. In the site representation, the Hamiltonian of the system can be written as follows:  ${\mathcal H}= E_1|1\rangle\langle 1|+E_2|2\rangle\langle 2|+(1/2)(V_{12}|1\rangle\langle 2|+h.c.)  + \xi(t) (\lambda_1|1\rangle\langle 1|+\lambda_2|2\rangle\langle 2|)$.
 We also assume that noise is produced by the stationary random telegraph process (RTP) with:
$\langle \xi(t)\rangle =0$,
 $\langle \xi(t)\xi(t')\rangle = \chi(t-t')$,
where, $\chi(t-t') = \sigma^2 e^{-2\gamma |t-t'|}$, is the correlation function; $\sigma$, $2\gamma$, and $\lambda_{1,2}$, are the amplitude of noise, the decay rate of the correlation function, and the interaction constants with noise, correspondingly.

 In the diagonal representation of the unperturbed  Hamiltonian, we obtain the total Hamiltonian for the effective TLS,
\begin{align}\label{Hqb2}
{\mathcal H}=\frac{\lambda_0}{2} I+\frac{1 }{2} \Omega\sigma_z + \frac{ 1 }
{2}D_{\lambda,z} \xi(t) \sigma_z \nonumber+ \\
 \frac{1 }{2} D_{\lambda,\bot}  \xi(t) (\cos\varphi\sigma_x + \sin\varphi\sigma_y),
\end{align}
where, $\sigma_{x,y,z}$, are the Pauli matrices, $\lambda_0 = E_1+ E_2 +(\lambda_1 + \lambda_2)\xi(t)$,  $\lambda= \lambda_1 - \lambda_2$, $\Omega =\sqrt{(E_1-E_2)^2 +|V_{12}|^2}$, $D_{\lambda,z} = \lambda\cos\theta$,  and
 $D_{\lambda,\bot} = \lambda\sin\theta$.  We set: $\cos\theta =(E_1-E_2) /\Omega$.

The dynamics of a TLS is described by two rates: the longitudinal relaxation rate, $\Gamma_1 = T_1^{-1}$, and the transverse relaxation rate, $\Gamma_2 = T_2^{-1}$. When the noise is weak, and the condition, $\tau_c \ll T_1,T_2 $, is satisfied  (where $\tau_c =1/(2\gamma)$ is the correlation time of the noise fluctuations), one can apply Bloch-Redfield (BR) theory \cite{BF1,Rag}. 
In BR theory, the transverse relaxation rate, $\Gamma_2 =   \Gamma_\varphi+\Gamma_1/2$, where   $\Gamma_\varphi$  is the so-called ``dephasing'' rate. In terms of the  spectral density of noise,
$S(\omega)$, these rates are defined as follows \cite{ICJM}:
$\Gamma_1 =  \pi D^2_{\lambda,\bot} S(\Omega)$, 
$\Gamma_\varphi =  \pi D^2_{\lambda,z} S(0)$. 
 Using the spectral density of RTP, $S(\omega)=2\gamma\sigma^2/\pi(4\gamma^2+\omega^2)$,  we obtain the relaxation and dephasing rates provided by BR theory:
$\Gamma_1 ={ 2\gamma v^2}\sin^2\theta/({4\gamma^2 
+ \Omega^2})$, $ \Gamma_\varphi =({ v^2}/{2\gamma})\cos ^2\theta$, where
 the renormalized interaction constant, $v=\lambda \sigma$, is introduced.
 
To study the quantum decoherence, we present the density matrix as, $ \rho(t) =(I+\mathbf n(t)\cdot\boldsymbol \sigma)/2$, where, $\mathbf n(t)=\rm Tr(\rho(t)\boldsymbol \sigma )$, is the Bloch vector.  Instead of the Liouville-von Neumann equation for the density matrix, $i\hbar \dot \rho = [{\mathcal H},\rho]$, it is convenient to employ the equation of motion for the Bloch vector (we set $\hbar=1$):
\begin{align}
  \frac{ d \mathbf n}{dt}= \mathbf\Omega \times \mathbf n + ({\xi(t)}/{\sigma})\boldsymbol\omega\times \mathbf n .
  \label{BlEq}
\end{align}
Here ${\boldsymbol\omega} =v(\sin\theta\cos\varphi,\sin\theta\sin\varphi,\cos\theta)$, and $ \mathbf\Omega =(0,0,\Omega)$. 

Using the differential formula for the RTP \cite{KV2}, 
\begin{align}
\Big(\frac{d}{dt} +2\gamma \Big)\langle  {\xi}(t)R[t;\xi(\tau) ]
\rangle   =\Big\langle  {\xi}(t)\frac{d}{dt}R[t;\xi(\tau)]
\Big\rangle, 
\end{align}
where,
$R[t;\xi(\tau)]$, is an arbitrary functional, we obtain from  Eq. (\ref{BlEq}) the closed system of differential equations:
\begin{align} \label{X1}
   \frac {d \langle\mathbf n\rangle}{dt}=& \mathbf\Omega \times\langle \mathbf n \rangle+ \boldsymbol\omega \times \langle{\mathbf n}^\xi\rangle, \\
    \frac {d \langle {\mathbf n}^\xi\rangle}{dt}=& \mathbf\Omega \times\langle  {\mathbf n}^\xi\rangle+ \boldsymbol\omega \times \langle{\mathbf n}\rangle - 2\gamma  \langle {\mathbf n}^\xi\rangle, 
    \label{X1a}
\end{align}
where, $\langle {\mathbf n}^\xi\rangle= \langle { \xi}(t)  \,{\mathbf n}\rangle /\sigma$. The average, $\langle...\rangle$, is taken over the RTP.

{ \em Homogeneous broadening.}    
To characterize a dimer,  we  introduce the dimensionless parameter, $\epsilon =|\tan\theta|=|V_{12}/(E_1-E_2)|$. 
When $\epsilon\ll 1$, we will call the dimer ``weakly coupled". In the opposite case, $\epsilon\gtrsim 1$, the dimer is called ``strongly coupled". 
 
 We first consider a weakly coupled dimer, so one can neglect the effects of relaxation. Introducing the complex vectors: $\langle m(t)\rangle = \langle n_x(t) \rangle + i \langle n_y(t) \rangle$ and   $\langle m^\xi(t) \rangle = \langle n^\xi_x(t) \rangle + i \langle n^\xi_y(t) \rangle$,  one can show that the solution of Eqs. (\ref{X1}) and (\ref{X1a})  can be written as,
\begin{align} \label{SL1}
\langle  n_z(t)\rangle =&\langle  n_z(0)\rangle, \quad \langle  n^\xi_z(t)\rangle =0,\\
\langle  m(t)\rangle =& e^{i\Omega t} \Phi(t) \langle  m(0)\rangle,\\
\langle  m^\xi(t) \rangle =&-\frac{i e^{i\Omega t}}{v\cos\theta}\frac{d\Phi(t)}{dt} \langle  m(0) \rangle.
\label{SL1a}
\end{align}
Here we denote by $\Phi(t)$  the generating functional  of the RTP \cite{KV2}.  For  the FI decay,  it is given by \cite{BGA,GABS,NB1},
\begin{align}
\Phi^f(t)= e^{-\gamma t}\Big(\frac{1}{\mu}\sinh({\gamma \mu t}) +\cosh({\gamma\mu t})\Big),
\label{FD_1}
\end{align}
where, $\mu = \sqrt{1-({ v}\cos\theta/\gamma)^2}$. 

{\em FI decay.}  We call noise weak if the dimensionless parameter, $\eta=|v \cos\theta/\gamma|\ll 1$. As it follows from Eq.  (\ref{FD_1}), for  weak noise, the decay rate of the non-diagonal averaged density matrix element (which characterizes the decoherence) coincides with $\Gamma_\varphi$, provided by BR-theory.
 We call noise strong if $\eta\gtrsim 1$. In particular, when $ \eta> 1$, the parameter, $\mu$ in Eq.  (\ref{FD_1}) becomes imaginary, and the decay of the functional, $\Phi^f(t)$, is accompanied by oscillations with frequency, $\gamma |\mu|$.   
 
Below, we compare the analytical solutions (\ref{SL1}) --  (\ref{SL1a}),  when the transverse effective field (relaxation) is neglected, with the corresponding exact solutions obtained numerically from Eqs. (\ref{X1}) and ({\ref{X1a}). In numerical simulations, we put $\hbar=1$. All energy-dimensional parameters are measured in $\rm ps^{-1}$ ($1\rm ps^{-1}\approx 0.66meV$), and time is measured in $\rm ps$. 

 In Fig. \ref{Fid1s}, the strongly coupled dimer was considered ($\theta \approx 0.968$, $\epsilon\approx 1.45$).  One can see, that in spite of the noise is strong: $\eta\approx 1.13$ (red dashed curve) and  $\eta\approx 2.27$   (orange dashed curve), the approximate analytical solutions (shown by blue curves), are in a good agreement with the exact numerical solutions.  
In the inset, the noise amplitude is relatively strong ($v=20$), and the matrix element, $V_{12}$, of the Chl1 and Chl2 interaction,  is large: $\epsilon\approx 8.24$ (strongly coupled dimer). At the same time, the noise is weak: $\eta\approx 0.24$, and the BR approach works. However, one cannot neglect the contribution from the transverse field to the decoherence rate, $\Gamma_2=\Gamma_\varphi+\Gamma_1/2$. Indeed, in this case, $\Gamma_\varphi\approx 0.3$ and $\Gamma_1\approx 0.48$. That is why the approximate solution in the inset (blue curve) deviates significantly from the exact numerical solution (red dashed curve). 

Our numerical simulations show that the analytical solution, given by Eqs. (\ref{SL1}) --  (\ref{SL1a}), is in a good agreement with the exact numerical solution, up to the value of $\epsilon \approx 1.72
$ ($\theta \lesssim \pi/3$), in spite of the dimer is already strongly coupled.
\begin{figure}[tbh]
 		\scalebox{0.425}{\includegraphics{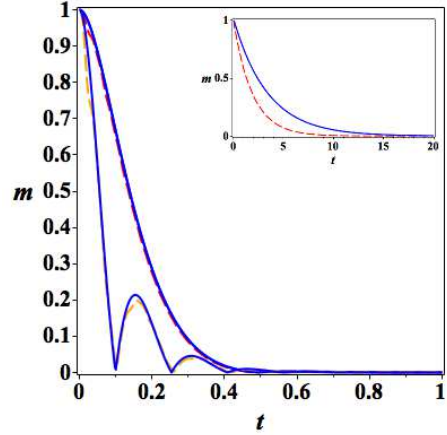}}
 	\caption{(Color online)  Time dependence (in ps) of $m=|\langle  m(t)\rangle |$, for the FI signal. Blue curves: analytical results. Dashed curves: exact solution. Parameters: $\Omega=127$, $\theta =  0.968$,  $\gamma=10$, $v=20$ (red curve), $v=40$ (orange curve). Inset: $\theta=1.45$, $v=20$.
 		\label{Fid1s}}
 \end{figure}

{\it Echo signals.} For simplicity, we assume that the $\pi/2$ and all spin echo $\pi$-pulses act practically instantaneously. The spin echo pulse, applied at the time $\tau$, rotates the wave function around the $x$-axis, by the angle $\pi$.  The corresponding analytical solution can be written as, 
\begin{align}
\Phi^e(t) = \left \{
\begin{array} {ll}
\Phi^f(t) ,& 0 < t < \tau, \\
\Phi_g^f(t) ,& t > \tau, 
\end{array}
\right.
\end{align}
where,
\begin{align}
\Phi_g^f(t)= &\Phi^f(t) 
 + e^{-\gamma t}\big(1-\frac{1}{\mu^2}\big)\big(\cosh(\gamma\mu (t-2\tau)) \nonumber \\
& - \cosh(\gamma\mu t)\big).
 \label{FI1a}
\end{align}

\begin{figure}[tbh]
 		\scalebox{0.35}{\includegraphics{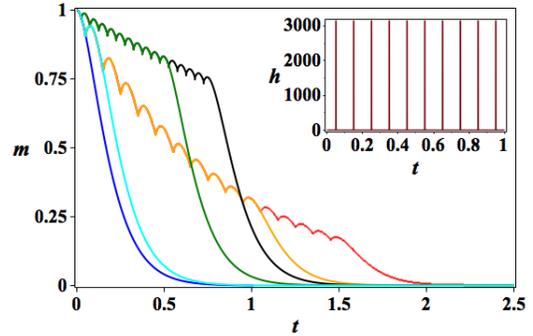}}
 	\caption{(Color online) Time dependence (in ps) of $m=|\langle  m(t)\rangle |$. The FI decay: blue curve. Echo signals: green curve:  $N=10$;  black curve:  $N=15$ ($\tau=0.025$); cyan curve: $N=1$; orange curve: $N=10$;  red curve:  $N=15$ ($\tau=0.05$). Parameters: $\Omega=10$, $v=10$,  $\gamma=10$, $\theta=0$. Inset: the sequence of echo $\pi$-pulses applied at times: $t=\tau, 2\tau,4\tau,\dots $. The duration of the echo $\pi$-pulse is: $\delta=1\rm fs$, and its height is: $h= 10^3\pi$.
 		\label{Echo1}}
 \end{figure}

\begin{figure}[tbh]
 		\scalebox{0.425}{\includegraphics{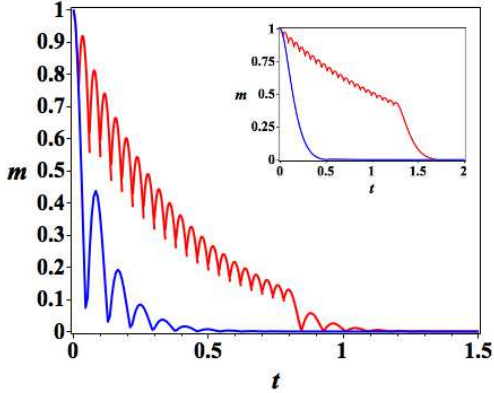}}
 	\caption{(Color online)  Time dependence (in ps) of $m=|\langle  m(t)\rangle |$.  The FI decay: blue curve. Echo signals: red curve  (number of echo pulses, $N=20$).  Parameters: $\Omega=150$, $v=40$,  $\gamma=10$, $\theta =0.165$, $\tau =0.02\, \rm ps$.  Inset: $\Omega=127$, $v=20$,  $\gamma=10$, $\theta =0.968$, $\tau =0.03$. 
 		\label{Echo2}}
 \end{figure}
 In Fig. \ref{Echo1},  the FI decay is shown for the transverse Bloch vector (blue curve). The  $N$ echo pulses are applied, partly restoring the FI signal, and reducing the {\it homogeneous broadening}. Green curve:  $N=10$;  black curve:  $N=15$ ($\tau=0.025$); cyan curve: $N=1$; orange curve: $N=10$;  red curve:  $N=15$ ($\tau=0.05$). 	Parameters: $\Omega=10$, $v=10$,  $\gamma=10$, $\theta=0$.  The inset corresponds the the sequence of echo $\pi$-pulses applied at times: $t=\tau, 2\tau,4\tau,\dots $. The duration of each echo pulse is: $\delta=1 \rm fs$, and its height is: $h= 10^3\pi$. In this case, the noise is strong: $\eta=1$, and the dimer is weakly coupled: $\epsilon=0$.  (There is no contribution to the dynamics from the transverse field.)

 In Fig. \ref{Echo2},  the time dependence (in ps) of the transverse Bloch vector, $m=|\langle  m(t)\rangle |$, is shown. The  FI decay corresponds to the blue curve. Echo signals: red curve  (number of echo pulses, $N=20$).  Parameters: $\Omega=150$, $v=40$,  $\gamma=10$, $\theta =0.165$, $\tau =0.02$.  This corresponds to a weakly coupled dimer ($\epsilon\approx 0.17$), and to  strong noise: $\eta\approx 3.9$. The FI decay exhibits oscillations with the period: $T=2\pi/\gamma|\mu|\approx0.16$.
  The inset corresponds to the choice of parameters: $\Omega=127$, $v=20$,  $\gamma=10$, $\theta=0.968$, $\tau =0.03$.  The noise is strong in this case, $\eta\approx 1.13$, and the  dimer is strongly coupled, $\epsilon\approx 1.45$. As one can see from Figs. \ref{Echo1} and \ref{Echo2}, the homogeneous broadening of the FI signal can be significantly improved by applying the spin echo pulses for strong noise and for both weakly and strongly coupled dimers.
 
{\em Simultaneous action of  the homogeneous and  inhomogeneous disorder.}  Here, in addition to dynamical fluctuations (noise, $\xi(t)$),  we consider an ensemble of TLSs (dimers) with  fluctuating parameters, $(\Omega, \theta,\varphi )$, due to the static disorder. It is well-known that this leads to the inhomogeneous broadening of the FI signal decay. In our numerical simulations we assumed the independent Gaussian disorder for parameters,  $(\Omega, \theta,\varphi )$. Our numerical simulations demonstrate that the main contribution from the static disorder, for a wide range of parameters, is due to the fluctuations of the frequency, $\Omega$. So, below we neglect the static fluctuations of both angles,  $\theta$ and $\varphi$. We assume a Gaussian distribution for the random parameter $\Omega$,  denoting the dispersion by $\sigma$. Note that the results can easily be extended by including the static fluctuations of angles, $\theta$ and $\varphi$, in the numerical solutions of the exact Eqs. (\ref{X1}) and (\ref{X1a}).

 In Fig. \ref{Fid1gr}, we compare the results for the decay of the FI signal for three values of the dispersion, $\sigma = 0,10,20$, of static fluctuations of the frequency, $\Omega$, and for the amplitudes of noise, $v=20,40$. The parameter, $\eta\approx 1.13$, so the noise is strong in this case. In the inset, oscillations are observed with the period, $T=2\pi/\gamma|\mu| $, due to the imaginary value of the parameter, $\mu$. 
\begin{figure}[tbh]
 		\scalebox{0.425}{\includegraphics{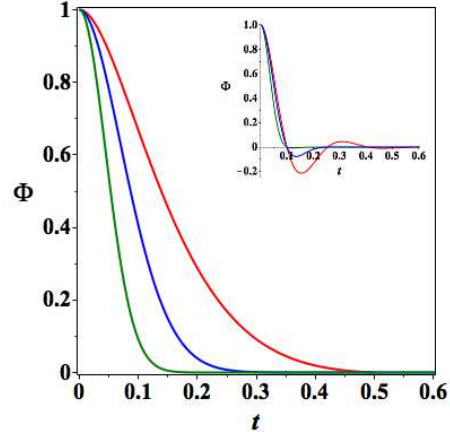}}
 	\caption{(Color online) Simultaneous action of homogeneous and  inhomogeneous broadening on the FI decay. The time dependence (in ps) of the  generating functional, $\Phi(t)$. Red curve: $\sigma =0$.  Blue curve: $\sigma= 10$. Green curve: $\sigma= 20$.  Parameters: $\Omega=127$, $\theta =  0.968$,  $\gamma=10$, $v=20$. Inset: $v=40$.   		\label{Fid1gr}}
 \end{figure}

The analytical solution, which  includes the contributions from both, the static disorder and the dynamical noise, and corresponds to the spin echo signal applied at the  time $\tau$, can be written as,
$\langle  m(t)\rangle =  \Phi^e_g(t) \langle  m(0)\rangle$,
where
\begin{align}
\Phi_g^e(t) = \left \{
\begin{array} {ll}
 e^{-\displaystyle \frac{\sigma^2 t^2}{2}}\Phi^f(t)  ,& \hspace{-0.75cm} 0 < t < \tau \\
~~~~\\
 e^{-\displaystyle \frac{\sigma^2 (t-2\tau)^2}{2}}\Phi_g^f(t),&t > \tau .
\end{array}
\right.
\end{align}
  \begin{figure}[tbh]
 		\scalebox{0.35}{\includegraphics{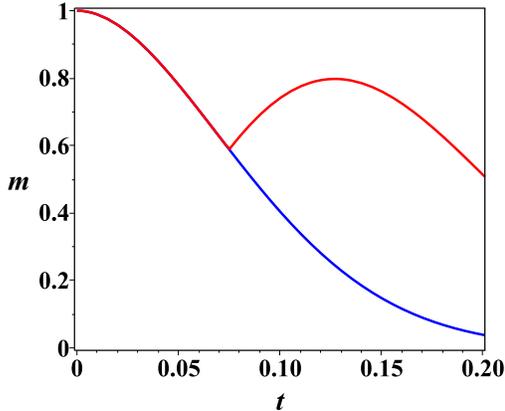}}
 	\caption{(Color online) The decay of the FI signal in the presence of both homogeneous and inhomogeneous broadening (blue curve), and the action of the echo signal (red curve). Time dependence (in ps) of  $m=|\langle  m(t)\rangle |$. Parameters: $\Omega=127$, $\theta =  0.968$,  $\gamma=10$, $v=20$, $\sigma =10$, $\tau =0.075$. 
 		\label{Fid1g}}
 \end{figure}

In Fig. \ref{Fid1g},  both static disorder of $\Omega$ and the dynamical noise, $\xi(t)$, are included. The decay of the FI signal is shown by the blue curve. For our chosen parameters, the FI signal decays in approximately $200 \rm fs$. The spin echo signal was applied at $\tau=75 \rm fs$. As one can see, the echo pulse restores significantly the FI decay (red curve). Note that for the parameters chosen in Fig. \ref{Fid1g}, both dimensionless decay factors coincide at the characteristic time of the FI decay, $t_*=200 \rm fs$: $\gamma t_*=\sigma^2t^2_*/2=2$. So, both homogeneous and inhomogeneous broadening are partly compensated in this case by the spin echo signal. \\

We presented the analytical and numerical results for spin echo pulses for the two-level systems (TLSs) -- chlorophyll-based dimers in bio-complexes, embedded in noisy protein environment.  We have shown that even strong dynamical broadening can be suppressed significantly by the spin echo pulses. This is important for many bio-applications at ambient conditions. We also demonstrated the restoration of the free induction decay signal by the spin echo pulses when both homogeneous and inhomogeneous broadening equally contribute to the free induction decay.
The application of the spin echo technique is especially useful for bio-systems with strong low-frequency noise, such as $1/f$ noise. Our approach can be easily generalized for this case, as was done in \cite{NB1}, by introducing the corresponding ensemble of the fluctuators. Also, many different sources of dynamical noises can be included in the presented here approach (as was done in \cite{NB2}).

\begin{acknowledgements}  
	
   This work was carried out under the  auspices of the National Nuclear Security Administration of the U.S. Department of Energy at Los Alamos National Laboratory under Contract No. DE-AC52-06NA25396. 
 A.I.N. acknowledges the support from the CONACyT, Grant No. 15349. G.P.B.  acknowledges the support from the LDRD program at LANL.
 
\end{acknowledgements}


\begin{thebibliography}{100}

\bibitem{BOOK}
M. Mohseni, Y. Omar, G.S. Engel, and M.B. Plenio (Eds.), {\em  Quantum Effects in Biology}, 
(Cambridge University Press, 2014).

\bibitem{AB}
R.J. Abraham, J. Fisher and P. Loftus, {\em Introduction to NMR spectroscopy}, (Wiley, Chichester, 1988). 

\bibitem{FL1}
H. Dong and G.R. Fleming, J. Phys. Chem. B, {\bf 118}, 8956 (2014).

\bibitem{ICJM}
G.~Ithier, E.~Collin, P.~Joyez, P.~J. Meeson, D.~Vion, D.~Esteve, F.~Chiarello, A.~Shnirman, Y.~Makhlin, J.~Schriefl, and G.~Sch\"on, {Phys. Rev. B}, {\bf 72}, 134519 (2005).

\bibitem{BGA}
J.~Bergli, Y.M. Galperin, and B.L. Altshuler,
 { New Journ.  Phys.} {\bf 11}, 025002 (2009).

\bibitem{GABS}
Y.M. Galperin, B.L. Altshuler, J.~Bergli, D.~Shantsev, and V.~Vinokur,
  {Phys. Rev. B}, {\bf 76}, 064531 (2007).
 
 \bibitem{NB1}
A.I. Nesterov and G.P. Berman, Phys. Rev. A, {\bf 85}, 052125 (2012).

\bibitem{BF1}
F.~Bloch.
{Phys. Rev.}, {\bf 105}, 1206 (1957).

\bibitem{Rag}
A.G. Redfield,
{ IBM J. Res. Dev.} {\bf 1}, 19 (1957).
 
\bibitem{KV2}
V.~Klyatskin,
 {\em Dynamics of Stochastic Systems,}
 (Elsevier, 2005).

\bibitem{NB2}
A.I. Nesterov and G.P. Berman, {\em The role of protein fluctuation correlations in electron transfer in photosynthetic complexes}, arXiv:1412.0512 (2014).
 

\end{thebibliography}
\end{document}